\documentclass[prb,twocolumn,showpacs,amsmath,amssymb,onlinecite,superscriptaddress]{revtex4-2}
\usepackage[english]{babel}
\usepackage{amsmath}
\usepackage{graphicx, nicefrac}
\usepackage{float}
\usepackage{siunitx}
\usepackage{comment}

\usepackage{dcolumn}
\usepackage{bm,graphicx}
\usepackage{etoolbox}

\usepackage[colorlinks]{hyperref}
\usepackage{url}
\usepackage[colorlinks]{hyperref}
\usepackage[normalem]{ulem}
\usepackage[dvipsnames,usenames]{color}
\usepackage{xcolor}
\usepackage[version=3]{mhchem} 

\usepackage{tcolorbox}

\usepackage{empheq}

\begin{document}

\title{Landau level spectroscopy in current solid state physics}

\author{Ana Akrap}
\email[]{aakrap@phy.hr}

\affiliation{Department of Physics, University of Zagreb, Croatia}

\author{Milan Orlita}
\affiliation{LNCMI, CNRS-UGA-UPS-INSA, 25, avenue des Martyrs, F-38042 Grenoble, France}
\affiliation{Institute of Physics, Charles University, CZ-12116 Prague, Czech Republic}
\email[]{milan.orlita@lncmi.cnrs.fr}  

\date{\today}

\begin{abstract}
Landau level spectroscopy plays an important role in modern condensed-matter physics. 
In this technique, electrons in a solid are subjected to quantizing magnetic fields and probed experimentally, often through optical methods. Direct and detailed insights into the electronic properties of crystalline materials are obtained, particularly the properties related to their band structure. Landau level spectroscopy enables the precise extraction of key parameters such as effective mass, carrier density, mobility, and band gap, and serves as a powerful tool for studying interactions between electrons and other quasiparticles in solids. Over its more than seventy-year history, Landau level spectroscopy has been applied mainly to semiconductors and semimetals. Today, its scope also includes graphene-based systems, surface and bulk states in topological materials, and other emergent systems with a narrow or vanishing band gap. In this work, we review the fundamentals of Landau level spectroscopy and illustrate them with selected examples from the literature.
\end{abstract}

\maketitle

\section{Introduction}

The electronic band structure is central to almost every physical property of a crystalline material. It determines which electronic states are allowed and occupied, what their energies are, whether electrons can move through the crystal, and how electrons in a solid interact with light, heat, and with each other. In this way, the electronic band structure governs the electrical, optical, magnetic, thermal, and even mechanical properties of crystalline materials. Knowledge of the electronic band structure is therefore essential for both fundamental science and virtually any conceivable application.

Modern condensed matter physics offers a number of experimental and theoretical methods on how to look at the electronic band structure. Today, the most visible and widely applied experimental techniques comprise angle-resolved photoemission spectroscopy (ARPES) and scanning tunneling microscopy and spectroscopy (STM and STS). These techniques are often applied alongside more traditional approaches, such as electrical transport and optical measurements, as well as thermodynamic probes such as calorimetry or magnetization. The experimental results are then commonly compared with the predictions obtained from theoretical and numerical \emph{ab initio} methods.

The choice of a technique or a combination of them depends on the type of information one seeks about a given crystal. For metallic solids, for example, when determining the density and effective mass of charge carriers, electron transport measurements are typically combined with infrared optics. To determine the band gap in semiconductors and insulators, optical spectroscopy in the appropriate spectral range is a suitable approach. To probe the electronic states at the surface of a solid, ARPES and STM/STS are among the methods of choice.

Experimental methods that explore the band structure of solids subjected to a magnetic field—in the regime of Landau quantization—play a prominent role. The phenomenon of Landau quantization, described in greater detail later, can be understood on the basis of high-school physics: a charged particle, for example an electron, undergoes periodic motion in a magnetic field, and its closed trajectory is known as a cyclotron orbit. Quantum mechanics then forces electrons to occupy only a discrete set of such orbits, which are called Landau levels. Applying a sufficiently strong magnetic field to a solid has an analogous effect and it causes the electronic states to become discretized.

The first experimental signatures of Landau quantization in solids were reported in 1930 by Shubnikov, de Haas, and van Alphen. In their celebrated studies of magnetotransport and magnetization in bismuth~\cite{Schubnikow30,deHaas30}, they found pronounced oscillations that were periodic in the reciprocal magnetic field ($1/B$). Soon thereafter, this was explained by Landau through an elegant theoretical framework ~\cite{LandauZfP30}, where he introduced what we nowadays know as Landau levels. Their degeneracy increases linearly in the magnetic field, $\zeta = eB/h$. Therefore, the Landau level occupation changes in a manner periodic to the reciprocal of the magnetic field. Today, such experiments belong to the most fundamental methods in solid-state physics~\cite{AshcroftMermin1976, Singleton01} and are commonly referred to as quantum oscillation measurements.  They allow us to map the shape of the Fermi surfaces of metals and semimetals with high precision by directly probing electrons or holes in the immediate vicinity of the Fermi level. This approach is often referred to as the fermiology of metals~\cite{AshcroftMermin1976}.

This review article focuses on an experimental technique that has been extensively used to study Landau-quantized solids since the 1950s: Landau level spectroscopy. This method probes the impact of Landau quantization on the optical response of solids, since light often couples to electronic excitations between different Landau levels. Monitoring these excitations in the reflectivity, absorption, or emission spectra of solids provides invaluable insight into their electronic structure. Among its key advantages, Landau level spectroscopy is, in our view, the most precise technique for determining the fundamental band-structure parameters, such as the effective mass and its anisotropy or the band-gap width. Unlike quantum oscillation measurements, it is a contactless method, and it also probes electronic states  away from the Fermi energy --- the initial and final Landau levels, and it can therefore equally be applied to insulating materials.

Despite its long history in solid state physics, Landau level spectroscopy has somewhat fallen into the shadow of more recently developed experimental techniques. Our goal is to present the strengths and limitations of this method and, in doing so, to help reestablish its relevance in modern solid state physics.

\begin{figure}[!t]
	\includegraphics[width=\linewidth]{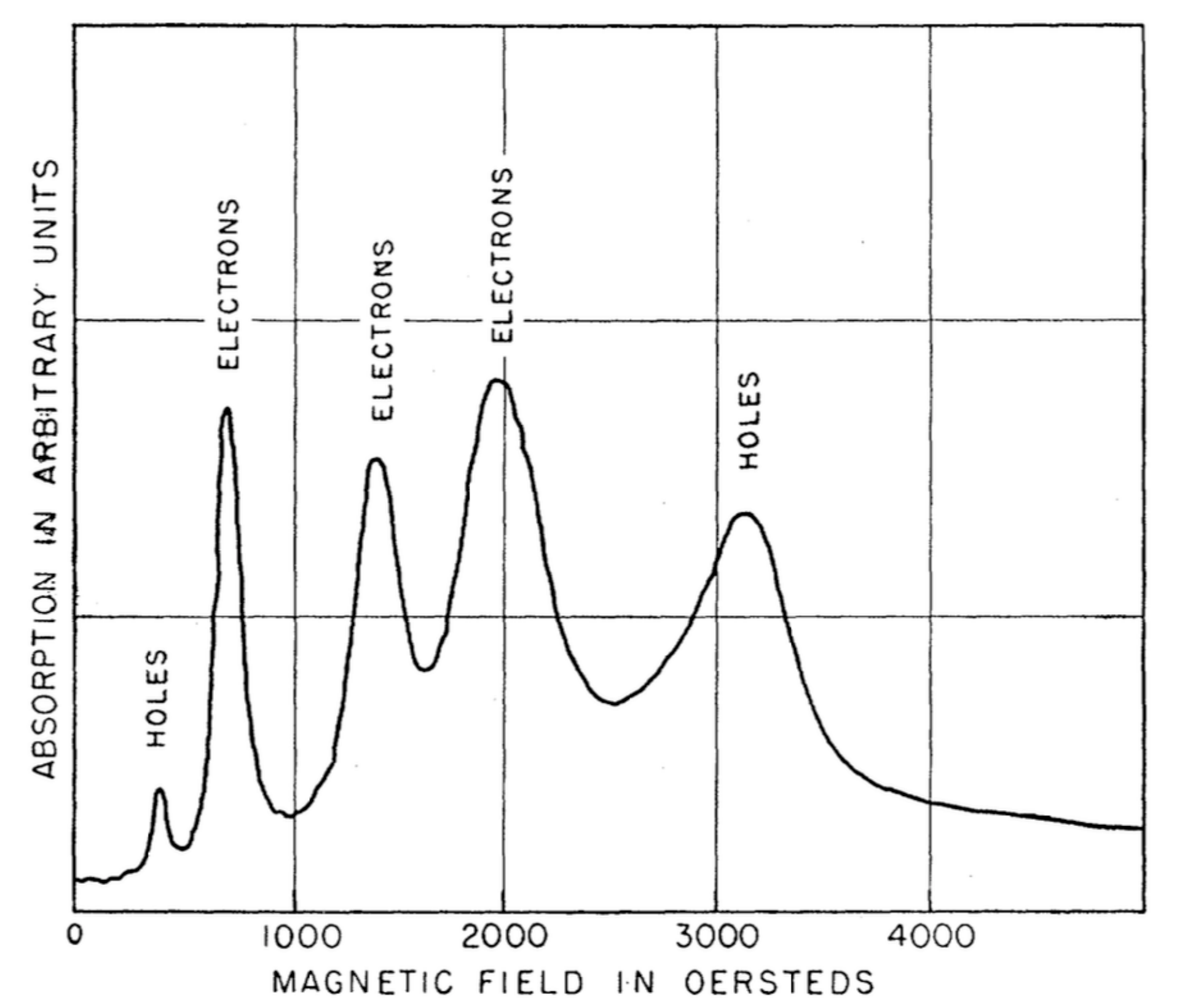}
	\caption{    
One of the first reported cyclotron resonance measurements, performed on germanium \cite{Dresselhaus:1955b}. In this experiment,  absorption of a monochromatic microwave (24 GHz) in germanium kept at low temperatures (4~K) was traced as a function of the applied magnetic field.
While we would expect to see only electrons or only holes, both kinds of carriers were seen in the experiments. The authors interpreted this as a result of their sample being ionized by microwaves.
Reprinted with permission from~\cite{Dresselhaus:1955b}, copyright (1955) by the American Physical
Society.}
	\label{fig:GeCR}
\end{figure}

\section{History of Landau level spectroscopy}

Landau level spectroscopy began in 1953.
This was closely linked to the first cyclotron resonance experiments conducted by Arthur Kip, Gene Dresselhaus, and Charles Kittel, who were the first to detect resonant microwave absorption in germanium crystals~\cite{Dresselhaus:1953}. They realized that what they were looking at were charged particles with an effective mass $m$ undergoing a cyclotron motion at a frequency $\omega_c=eB/m$, a formula many of us learned in high school. Although the effect could be explained entirely using classical physics, it opened the way to the development of optical spectroscopy in truly Landau-quantized solids.

This observation was far from obvious. It gave direct evidence that quasiparticles in solids truly existed and that they largely behaved like electrons but possessed an effective mass distinct from electrons in vacuum. At that time, the method for determining the effective mass from the temperature damping of quantum oscillations had not yet been fully developed~\cite{LifshitzSPJ54}. The observation of cyclotron resonance therefore played a key role in establishing the concept of quasiparticles in solids~\cite{CohenAIPCP2005}. 

\begin{figure}[b]
    \centering
    \includegraphics[width=0.99\linewidth]{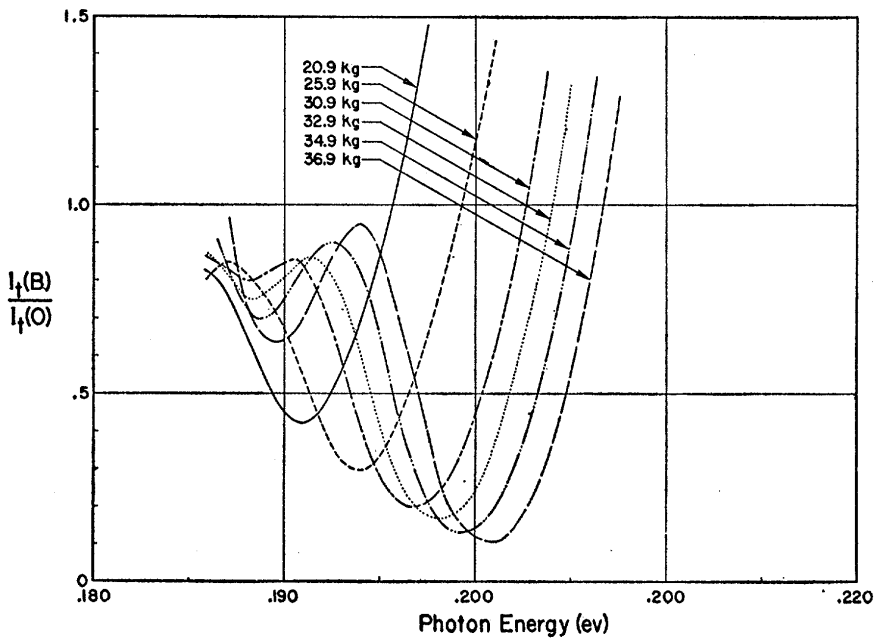}
    \caption{Modulation of mid-infrared transmission through a crystal of InSb,  induced by the magnetic field, was one of the very first optical experiments showing Landau quantization in a solid via interband optical excitations. The plot shows relative magneto-transmission for selected values of magnetic field. The dip corresponds to an interband inter-Landau level transition, which is a remarkable observation at such low magnetic fields ($2-4$~T) and at room temperature. Reprinted with permission from~\cite{ZwerdlingPR57}, copyright (1957) by the American Physical Society.}
    \label{fig:InSb}
\end{figure}

\begin{figure*}[t]
    \centering
    \includegraphics[width=0.65\linewidth]{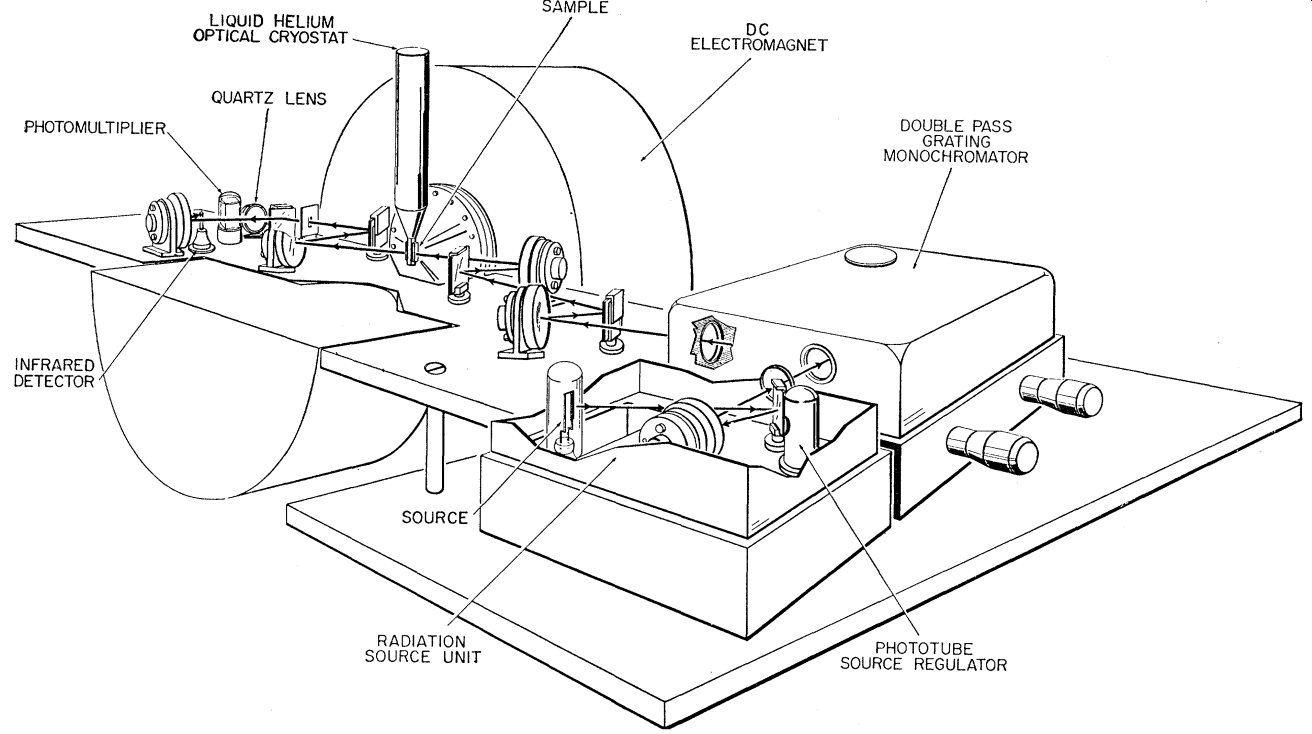}
    \caption{Cut-away drawing of an early experimental setup developed for Landau level spectroscopy of semiconductors in late 1950's, prior to Fourier transform spectroscopy was introduced during 1960s. It is composed of a radiation source, a grating monochromator, a cryostat and an electromagnet. This setup was used to study interband excitations in Landau-quantized germanium~\cite{ZwerdlingPR59}. Reprinted with permission from~\cite{ZwerdlingPR59}, copyright (1959) by the American Physical Society.}
    \label{fig:EarlySetup}
\end{figure*}

The pioneering measurements on germanium (see Fig.~\ref{classicalCR}) were soon followed by cyclotron-resonance experiments in many other crystals: silicon~\cite{Dresselhaus:1955b}, bismuth~\cite{GaltPR55}, InSb~\cite{Dresselhaus:1955a}, and graphite~\cite{GaltPR56}, to name a few. Experiments probing interband excitations in Landau-quantized materials appeared shortly thereafter; see, for example, studies performed on germanium and InSb~\cite{BursteinPR57,ZwerdlingPR59,ZwerdlingPR59}, illustrated in Figs.~\ref{fig:InSb} and
\ref{fig:EarlySetup}. These early works firmly established Landau level spectroscopy as a powerful method for investigating the electronic band structure of solids.

Landau level spectroscopy is in principle applicable to any crystalline material, 
including metals, semimetals, semiconductors, and insulators. In practice, however, its development has always been constrained by experimental limitations, most notably the availability of magnetic fields strong enough to induce well-resolved Landau levels. Therefore, in the 1960s Landau level spectroscopy evolved primarily as a tool for studying semimetals and semiconductors~\cite{Landwehr,RashbaSPU65, PalikRPP70,McCombePRL67}. Although initially applied to bulk materials, the technique was later extended to quantum wells and heterostructures, as described in an early review~\cite{AndoRMP82} and many experimental examples ~\cite{GuldnerPRL80,VoisinAPL81,HeitmanPRB86, GlaserPRB87,SingletonSS88,ShuvaevPRB17}. In semiconductors and semimetals, the electronic states close to the Fermi energy are often formed from the $s$ and $p$ states of the constituent atoms. The strong overlap of these atomic orbitals produces steep, highly dispersive electronic bands, which are characterized by a low density of states and small effective masses. Materials with light carriers are easily Landau-quantized when a magnetic field is applied. These properties favor the use of optical spectroscopic techniques operating in the terahertz and infrared spectral ranges.

Cyclotron-resonance absorption, using microwave techniques, has also been successfully observed in many conventional metals~\cite{KipPR61,GrimesPR63} and even in strongly correlated systems~\cite{KimataPRL11}. Still, the overall impact of Landau level spectroscopy in metals remains limited in comparison to its importance for semiconductors and semimetals. Although Landau quantization is clearly possible in metals, their electronic bands are generally much flatter because they originate from weakly overlapping $d$ or $f$ states. These flatter bands imply a much higher density of states and relatively higher masses. Consequently, achieving Landau quantization in metals typically requires stronger magnetic fields.

Over the years, Landau level spectroscopy developed far beyond a method simply probing the shape and distance of electronic bands. Optical spectroscopy of Landau-quantized electrons also became a direct tool for studying the interaction of electrons with other quasi-particles in a crystal. The interaction of electrons with phonons is often manifested in the optical response of Landau quantized solids, allowing us to explore polaronic effects~\cite{WuPRB86,FaugerasPRL04}. Moreover, the so-called magnon-phonon resonance~\cite{NeumannNL15,FaugerasJRS18} may emerge when the energy of an inter-Landau level excitation resonates with a certain phonon mode. The interaction of cyclotron resonance excitations with (confined) plasmons is another intensely explored area in Landau level spectroscopy~\cite{BatkePRL85,BandurinNP22}.

The past quarter century has brought new possibilities and challenges for Landau level spectroscopy. First, the appearance of truly two-dimensional materials, with their thickness down to a single atomic layer, created a unique opportunity. A number of these materials are characterized by a small or even vanishing band gap. Graphene and its few-layer stacks are a prime example. Second, immense progress has been made in our understanding of electronic band structure from the viewpoint of topology. This allowed us to look at a number of previously known materials and find formerly ignored or even unknown band structure features. The surface states of topological insulators or three-dimensional Dirac and Weyl semimetals represent perfect examples. These gapless phases represent a natural target for Landau level spectroscopy studies.

\section{Generating high magnetic fields}

A key condition to obtain Landau quantization is having a sufficiently high magnetic field. High fields are needed to overcome the scattering of charge carriers caused by impurities and other defects that are always present in all solids and prevent Landau levels from forming. This may be expressed in terms of electron mobility $\mu$, which allows us to estimate the lowest magnetic field needed for a crystalline solid to enter the Landau quantization regime, $B>1/\mu$. Landau level spectroscopy often targets novel materials, at an early stage of research, and therefore with a limited electronic quality linked to a lower electron mobility. This is why Landau level spectroscopy is strongly tied to the state-of-the-art high magnetic field technology~\cite{BattestiPR18}.

Developments of powerful water-cooled resistive electromagnets~\cite{BattestiPR18} have been primarily centered in several high magnetic field laboratories around the world. Starting with the 1930s, pioneering work in this respect has been made 
in Berkeley by William F. Giauque, and in Cambridge, Boston, by Francis Bitter~\cite{HerlachPB95,BitterRSI39}. Later, since the 1970s, important progress was achieved thanks to the French-German collaboration in the joint CNRS-Max-Planck-institute laboratory in Grenoble~\cite{PauthenetEPN74}. While in 1950s the generated magnetic fields rarely exceeded 10~T, nowadays, $dc$ magnetic fields near or even above 40~T are available in Tallahassee, Hefei, Grenoble and Nijmegen. Moreover, magnetic fields reaching the strengths of tens and even hundreds of teslas are accessible in laboratories of pulsed magnetic fields in Toulouse, Los Alamos, Wuhan, and Tokyo.

In addition to powerful resistive electromagnets that back in the day existed only in a select few laboratories, superconducting coils started to emerge in the 1960s~\cite{SampsonSA67}. Already in the 1970s, superconducting magnets surpassed the resistive ones in practicality and stability for the generation of moderate magnetic fields. Such magnets are now routinely used in many solid-state laboratories, capable of delivering magnetic fields in excess of 20~T.  Many of them are regularly used for Landau level spectroscopy.

\begin{figure}[b]
    \centering
    \includegraphics[width=0.95\linewidth]{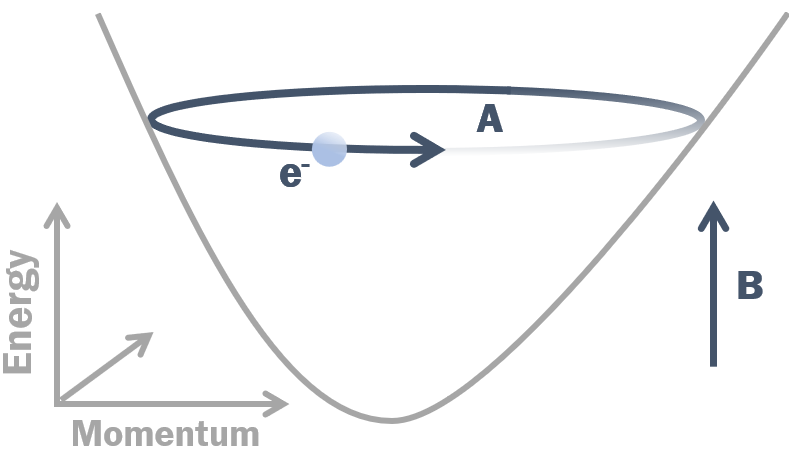}
    \caption{Cyclotron motion of an electron in an arbitrary band. The electron encircles an area $A$, whose energy dependence defines the effective cyclotron mass, see Eq.~\ref{classicalCRmass}.}
    \label{fig:A}
\end{figure}

\section{Cyclotron motion and resonance}

Before delving into details of Landau quantization and related spectroscopy techniques, let us briefly discuss the impact of the magnetic field on charged particles
in vacuum, in the scope of classical physics. 

The classical cyclotron motion  refers to the kinematics of a free charged particle subjected to a uniform magnetic field $\mathbf{B}$. In the plane perpendicular to 
$\mathbf{B}$, such motion is perfectly circular, and the associated frequency---the so-called cyclotron frequency---is given by the familiar expression:
\begin{equation}
\omega_c=\frac{|q|B}{m},
\label{classicalCR}
    \end{equation}
where $q$ is the charge of the particle and $m$ represents its mass. When electrons undergoing such a cyclotron motion are hit by light with a frequency $\omega$ that approaches $\omega_c$, the radiation is resonantly absorbed. This phenomenon is called \emph{cyclotron resonance absorption}.

Naturally, free particles in solids---electrons or holes---also undergo such cyclotron motion when a magnetic field is applied. However, the complexity of electronic bands in crystals with various space symmetries means that such cyclotron motion is not necessarily circular, neither in direct space, nor in reciprocal space. Hence, it may be surprising that the cyclotron frequency can still be expressed by (\ref{classicalCR}), if we take the appropriate mass, called the cyclotron mass.

The cyclotron mass $m_c$ is a quantity defined for an electron or hole circulating along a cyclotron orbit, in any kind of electronic band~\cite{AshcroftMermin1976}:
\begin{equation}
m_c=\frac{\hbar^2}{2\pi}\frac{dA}{dE},
\label{classicalCRmass}
\end{equation}
where $A$ is the energy-dependent area enclosed by the trajectory in the reciprocal space, as illustrated in Fig.~\ref{fig:A}. Although it may be difficult to evaluate the cyclotron mass for a complex band structure, simple results can be obtained for ideal parabolic and conical bands. These are, of course, two very specific cases, but they are often encountered in solids. Moreover, parabolic and conical bands represent extremely useful starting points that allow us to understand the response of more complex band structure features.

In a parabolic band, the energy disperses parabolically with momentum, $E=\hbar^2k^2/(2m)$. Parabolic bands naturally appear near the band edges, either as local or global extrema of electronic bands. There, the cyclotron mass (\ref{classicalCRmass}) becomes equal to the band mass, $m_c=m$, which is a result that follows our intuition trained by classical physics. Conversely, at a crossing point of two electronic bands, one finds a conical feature. The energy dispersion then becomes linear in momentum, $E=\hbar v |k|$. The cyclotron mass then becomes energy-dependent, $m_c=E/v^2$: it increases linearly with the distance from the crossing point and scales inversely to the square of the velocity parameter $v$. This result can be thought of in terms of Einstein's energy-mass relation.

\begin{figure}[t]
	\includegraphics[width=0.85\linewidth]{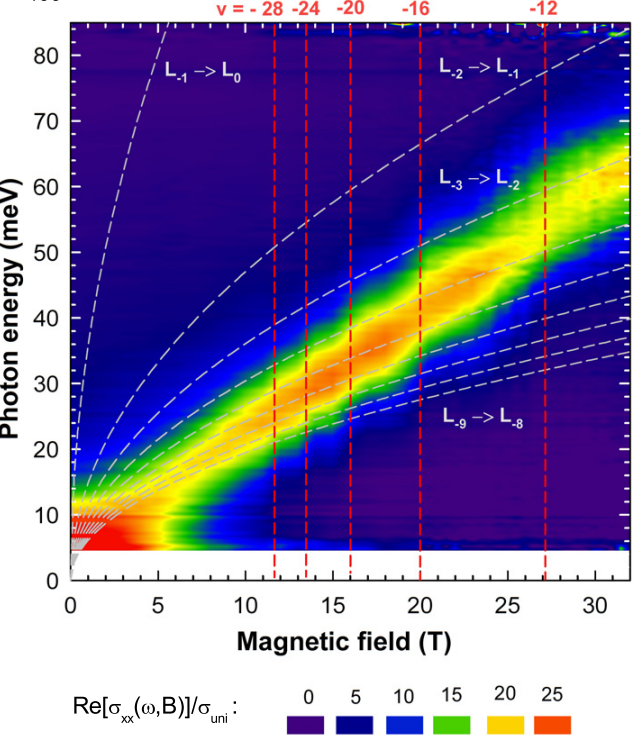}
	\caption{Magneto-absorbance (the real part of optical conductivity) 
    of highly doped graphene measured at low temperatures. The observed transition is linear in $B$, and it is a quasi-classical cyclotron resonance. The intensity and width of the mode is slightly modulated with $B$, which is an indication that the regime of Landau quantization is approached. In this regime, the $\sqrt{B}$ scaled transitions are shown in gray dashed lines, and particular filling factors indicated by red dashed lines; see \cite{OrlitaNJP12} for details.}
	\label{ClassicalGraphene}
\end{figure}

The classical treatment of cyclotron resonance, which led to formula (\ref{classicalCR}), has only limited validity. In sufficiently high magnetic fields, a quantum-mechanical approach is needed to describe how the Landau quantization impacts the optical response of a solid. This will be discussed in the next section. In contrast, in the regime of low magnetic fields,  the carrier scattering---ubiquitous in all solids---dominates the response. To give meaning to the terms \emph{high} and \emph{low} magnetic fields, we need to compare the cyclotron frequency $\omega_c$ to the scattering rate of charge carriers $\tau^{-1}$. The adjectives \emph{high} and \emph{low} then refer to the situation when $\omega_c \tau\gg1$ and $\omega_c \tau\ll1$, respectively. The strength of scattering therefore decides whether the quasi-particles undergo many turns on a cyclotron orbit before being scattered, or vice versa, no orbit is closed at all. The cyclotron resonance that follows the classical formula (\ref{classicalCR}) is observed in the intermediate regime, at the onset of closed cyclotron orbits, when $\omega_c \tau \sim 1$ and the scattering rate and the cyclotron energy are comparable.

There are many experimental examples of cyclotron resonance in the classical, or better said, in the quasi-classical limit discussed in the literature. In systems with strictly parabolic bands, there is no notable difference between the quasi-classical and fully quantum regimes~\cite{PalikRPP70}. As we shall see, the response in both cases consists of a single mode at the classically defined cyclotron frequency~(\ref{classicalCR}). In contrast, materials with non-parabolic bands are much more interesting from this perspective. An excellent example is graphene, with its widely extending conical bands. There, the quasi-classical cyclotron resonance, strictly linear in $B$, stands in stark contrast to the $\sqrt{B}$ dependence which is a hallmark of Landau-quantized conical bands, see e.g., Refs.~\cite{CrasseeNP10,WitowskiPRB10,OrlitaNJP12,GoerbigRMP11}. An example of cyclotron resonance in graphene is shown in Fig.~\ref{ClassicalGraphene}. The quasi-classical regime is imposed by the rather large doping in the explored sample, which shifts the Fermi energy far from the apex of the cone and leads to a rather large effective cyclotron mass $m_c=E_F/c^2$. At the same time, the observed modulation of the integral intensity and mode width is connected with the onset of Landau quantization. We will discuss this in the next section.

\begin{figure}
    \centering
    \includegraphics[width=0.95\linewidth]{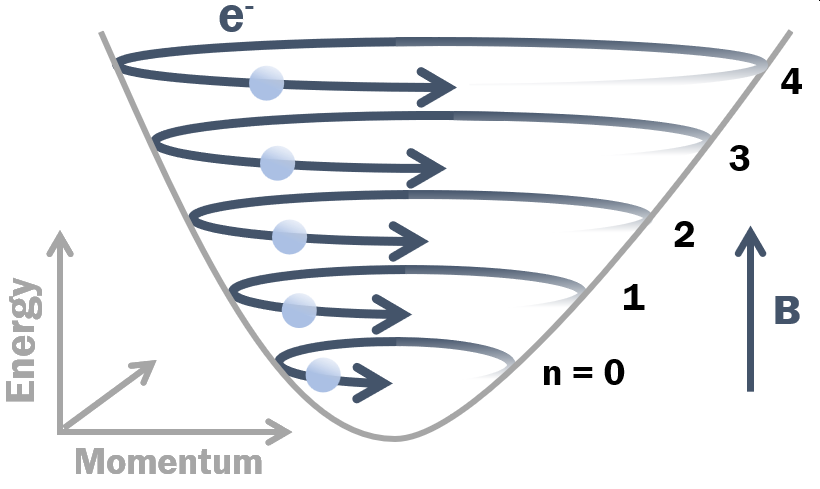}
    \caption{Schematic view of cyclotron orbits (Landau levels) allowed by Sommerfeld quantization rule.}
    \label{fig:Sommerfeld}
\end{figure}

\begin{figure*}[t]
	\includegraphics[width=0.8\linewidth]{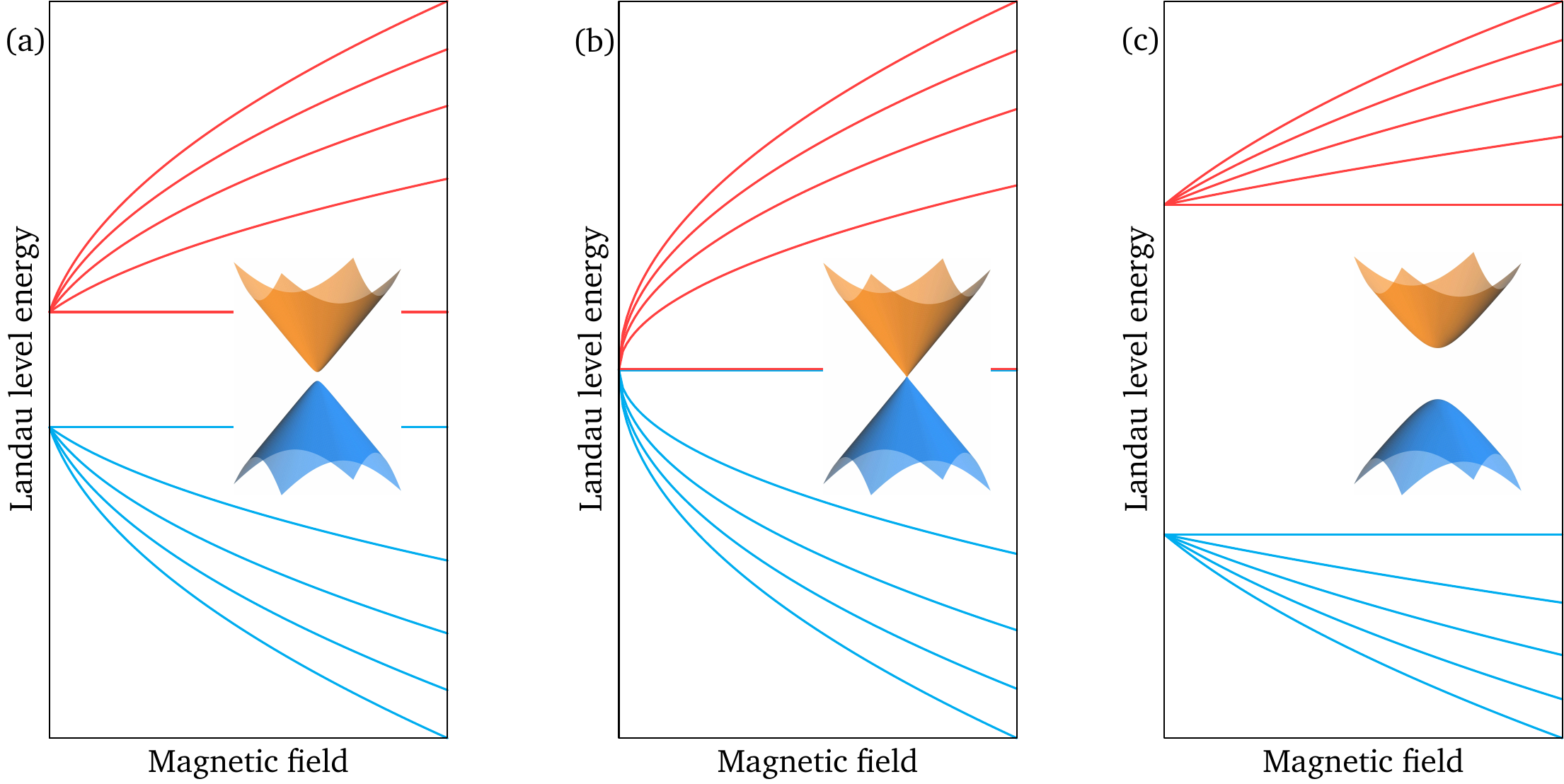}
	\caption{    
    Schematics of Landau levels for band structure described by the massive Dirac electron model, calculated using expression (\ref{eq:massD}).
    (a) Massive Dirac model, with a small gap parameter $\Delta$.
    (b) Conical band with a vanishing gap parameter, $\Delta = 0$, such as in graphene.
    (c) A large gap parameter $\Delta$ gives rise to the limit of parabolic bands.
    The field dependence of Landau levels differs profoundly. The $\sqrt{B}$ dependence, shown in (b), is a hallmark of conical bands, while the linear in $B$ dependence is typical of systems with a parabolic dispersion. 
    In each case, the zero-field extrapolation of interband excitations between Landau levels gives a reliable estimate of the band gap.}
	\label{fig1}
\end{figure*}

\section{Landau quantization of electrons in solids}

Applying a sufficiently strong magnetic field ($\omega_c \tau\gg 1$) leads to the formation of Landau levels. The field-induced quantization transforms the originally continuous density of states into a set of discrete or quasi-discrete energy levels, see Fig.~\ref{fig:Sommerfeld}. This can be understood through Sommerfeld quantization~\cite{OnsagerLEDPMJC52} --- quantum mechanics requires that the length of a closed orbit be a multiple of the de Broglie wavelength. This transformation has a profound impact on a number of physical properties. For example, the electrical conductivity and magnetization become strongly modulated, resulting in the well-known Shubnikov-de Haas or de Haas-van Alphen effects~\cite{Schubnikow30,deHaas30}. Landau quantization also manifests itself in optical experiments. Such experiments are referred to as the Landau level spectroscopy technique. We will discuss this in detail now.

Under practically attainable magnetic field strengths, Landau quantization typically occurs only in specific regions of the Brillouin zone. This is because the effective cyclotron mass and scattering time vary across the Brillouin zone and also depend on energy. As the magnetic field increases, certain regions of the Brillouin zone and the corresponding energy ranges enter the Landau quantization regime earlier than others. These regions are generally characterized by a low density of states, and the band energies are often close to the Fermi level. In most cases, they correspond to minima of the lowest conduction band, or maxima of the highest valence band, or lie near those points in the reciprocal space where two or more bands intersect and touch.

To introduce how Landau levels are formed in such regions, let us first recall the textbook model of massive Dirac electrons. In quantum electrodynamics, this model describes the behavior of massive fermions~\cite{Landau-Lifshitz71}. In the context of solid-state physics, this model refers to two mutually coupled electronic bands
whose extrema are located at the same place in the Brillouin zone and typically close to the Fermi energy. The usually empty conduction ($c$) and occupied valence ($v$) bands have characteristic relativistic-like hyperbolic profiles, with a full electron-hole symmetry: 
\begin{equation}\label{eq:MassDiracDisp}
E_{c}=-E_{v}=\sqrt{\Delta^2+v^2\hbar^2k^2}.    
\end{equation}
At low momenta, the bands are approximately parabolic, separated by the band gap of $2\Delta$ and characterized by the band-edge mass of $\Delta/v^2$. For large momenta, the electronic bands asymptotically approach the linear-in-momentum dispersion typical of massless charge carriers, $E=v\hbar|k|$.

This model was used as early as the 1960s to describe electronic bands in semimetals and direct-gap semiconductors; see the review by Zawadzki \cite{ZawadzkiJPCM17} and the references therein. The relevance of the massive Dirac model has increased with the recent emergence of graphene and other 2D materials, as well as topological matter~\cite{GoerbigEPL14}. In many cases, this model allows us to understand the magneto-optical response, or at the very least it facilitates a qualitative assessment of the experimental data.

In magnetic fields, the motion of electrons in the conduction and valence bands becomes quantized into Landau levels. From a semiclassical point of view~\cite{OnsagerLEDPMJC52}, this quantization means selecting only specific cyclotron orbits in both real and reciprocal space. In real space, one may imagine that the orbits allowed by quantum mechanics are chosen by placing a constraint on the magnetic flux $\Phi$ passing through the area $S$ enclosed by the electron. 
The quantization condition requires this flux to be an integer multiple of the magnetic flux quantum: $\Phi =B\cdot S = n h/e$. An equivalent condition can be formulated in reciprocal space. There, the semiclassical quantization rule selects discrete cyclotron orbits whose radius is $k_n = \sqrt{2enB/\hbar}$, labeled by the integer $n$. This condition allows us to use Eq.~(\ref{eq:MassDiracDisp}) and get the Landau level spectrum of massive Dirac electrons:
\begin{equation}\label{eq:massD}
E^n_{c}=-E^n_{v}=\sqrt{\Delta^2+v^22e\hbar nB}.    
\end{equation}
In this specific case of massive Dirac electrons, we have found the correct form of Landau levels simply by using quasi-classical arguments. In  general, however, finding the right Landau level spectrum is often a complex quantum-mechanical problem~\cite{GoerbigEPL14,FuchsSP18}. To simplify the Eq.~(\ref{eq:massD}), we neglected the motion of electrons along the direction of the applied magnetic field,  relevant only for 3D materials.

The massive Dirac electron model is frequently used to reproduce an observed series of inter-Landau level resonances, notably in narrow gap semiconductors. However, there are two limiting cases of this model that deserve our special attention: a large and, in contrast, vanishing gap parameter $\Delta$.
For a vanishing gap parameter, the two branches of the hyperbolic dispersion  transform into two conical bands meeting at their apexes, and the Landau level spectrum gets the characteristic $\sqrt{B}$ dependence:
\begin{equation}
E^n_{c}=-E^n_{v}=v\sqrt{2e\hbar nB}.    
\end{equation}
This $\sqrt{B}$ dependence is an unmistakable signature  of massless particles---well-known in the solid state community since the birth of graphene. Still, let us note that not every massless electron in a solid has to be described by the Dirac Hamiltonian. There are other models frequently applied to various materials, such as Kane~\cite{KaneJPCS57,Orlita2014} or Rashba~\cite{RashbaSPST60,BordacsPRL13} models, which may give rise to conical bands, and which then similarly lead to Landau levels with the $\sqrt{B}$ dependence
\footnote{Diracness is not a property of dispersion, but a property of the wave function.}.

The opposite limit, that of a large gap parameter $\Delta$, describes the situation when the relevant energy scale of electrons or holes---counted from the corresponding band-edge---is small in relation to $2\Delta$. The hyperbolic profile of the dispersion can then be well approximated with a parabola, and the Landau level spectrum becomes exactly linear in the applied magnetic field:
\begin{equation}
E^n_{c}=-E^n_{v}=\Delta+\hbar\omega_c(n+1/2)+g\mu_B B. 
\label{Schrodinger}
\end{equation}
Here $\omega_c$ is the cyclotron frequency and $g$ is the gyromagnetic factor. Sticking strictly to the massive Dirac model, we always get $\omega_c=eBv^2/\Delta$ and $g=2$. Nevertheless, one should consider a more general case, with arbitrary masses $m_e$ and $m_h$, as well as different $g$ factors for electrons and holes, $g_e$ and $g_h$. These generalizations allow us to extend this model to understand the response of various semiconducting compounds.

\section{Optical excitations in Landau quantized materials}

Having the electronic band structure of a solid---or more realistically, only parts of it---quantized into Landau levels, we may use a number of experimental methods to probe this band structure. In its modern use, the term \textit{Landau level spectroscopy} refers to any experimental method that allows us to visualize the energy spectrum of Landau-quantized electrons. This also includes gate-tunable electron transport or STM/STS in magnetic fields, see, e.g.,  \cite{ButtnerNP11,LiScience17} and \cite{LiPRL09,HanaguriPRB10}, respectively. Historically, Landau level spectroscopy used to be much more narrowly connected with optical spectroscopy techniques, and employed to study how radiation is emitted, absorbed, reflected, or scattered by a Landau-quantized solid.

In this review, for the sake of simplicity, we will focus only on those magneto-optical effects that can be described using linear response theory~\footnote{This excludes, among other, the large body of work with Raman scattering and photoluminescence studies of Landau-quantized solids, see, e.g., \cite{Pellegrini07,KossackiPRB11,FaugerasJRS18,PotemskiPB98,RomestainPRL80}}. The optical response, at a given frequency of radiation $\omega$ and an applied magnetic field $B$, is then fully encoded in the complex optical conductivity, $\sigma(\omega,B)$. For example, the absorption coefficient---due to electrons promoted by light between pairs of Landau levels---is in the simplest approach directly proportional to the real part of the optical conductivity, 
$\mathrm{Re}\{\sigma(\omega,B)\}$. Depending on the preferred experimental geometry, one may express the optical conductivity using the basis of Cartesian coordinates or circular polarizations of the electric field. 

\begin{figure*}[!t]
        \centering
    \includegraphics[width=0.8\linewidth]{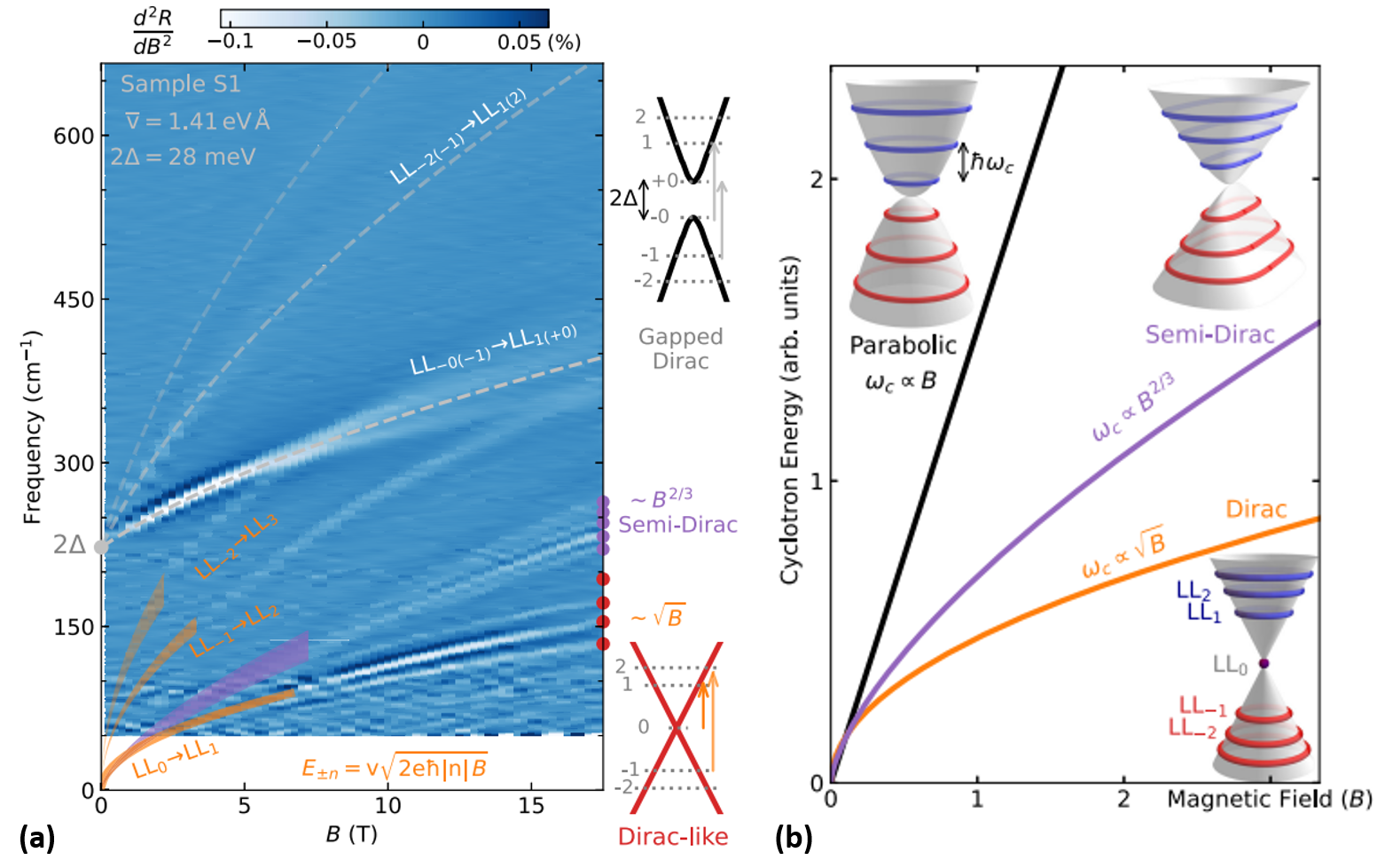}
    \caption{(a) Magneto-reflectance of ZrSiS with three distinct series of inter-Landau-level excitations identified. (b) Characteristic dependence of Landau levels in conical, parabolic (massive Dirac) and combined, semi-Dirac bands. Adapted from Ref.~\cite{ShaoPRX24}.}
    \label{fig:Semidirac}
\end{figure*}

Optical conductivity can be calculated employing the Kubo formula ~\cite{KuboJPSJ57}. This formula can be further simplified in the Kubo-Greenwood approach~\cite{GreenwoodPSC58,MoseleyAJP78} which neglects interactions among electrons and is basically equivalent to the Fermi golden rule. The real part of optical conductivity may then be schematically disentangled:
\begin{widetext}
\begin{multline}
\mathrm{Re} \{\sigma(\omega,B)\} \propto \frac{1}{\omega} \times\boxed{\substack{\mathrm{Landau\,level} \\ \mathrm{degeneracy}}} \times
                             \boxed{\substack{\mathrm{Valley} \\ \mathrm{degeneracy}}} \times
                            \boxed{\substack{\mathrm{Spin} \\ \mathrm{degeneracy}}} \times 
\sum_{n,m} \underbrace{{\cal{M}}_{m,n}}_{\substack{\mathrm{Selection} \\ \mathrm{rules}}} \times 
\underbrace{(f_n-f_m)}_{\substack{\mathrm{Occupation} \\ \mathrm{factor}}} \times 
\underbrace{\delta(E_m-E_n-\hbar\omega)}_{\substack{\mathrm{Energy} \\ \mathrm{conservation}}} 
\label{Kubo}
\end{multline}
\end{widetext}
In this way, the overall absorption of light is represented as a weighted sum of individual optical excitations that bring electrons from the Landau level $m$ to $n$. Each of these excitations has an intensity governed by several factors. Let us now describe each factor and, when appropriate, provide an illustrative experimental example from the literature.

\subsection*{Overall strength} 

Three factors are common to all excitations. The Landau level degeneracy, $\zeta=eB/h$, is a multiplicative factor that counts the states in each Landau level, empty or occupied, 
which may participate in light absorption. The second factor comes from the valley degeneracy, and it has the conventional meaning. Typically, it counts the non-equivalent band extrema in the reciprocal space, but it may change as a function of $B$ due to magnetic breakdown. The spin degeneracy factor is equal to 2 in systems with effectively missing spin-orbit coupling, such as graphene, but it is reduced to 1 in materials where this coupling cannot be neglected. For those materials, the spin and orbital motion cannot be separated, and the spin-degree of freedom is inherently included in the Landau level spectrum.

\subsection*{Position and shape of inter-Landau level excitations} 

Three factors are specific to each $n \rightarrow m$ inter-Landau-level excitation.
The $\delta$-function represents the energy conservation in the absorption process and matches the energy separation of two Landau levels with the energy of the incoming photon. In the simplified formula (\ref{Kubo}), this gives rise to an infinitely sharp excitation in the spectrum, at the energy $E_m-E_n$. The characteristic $B$-dependence of such an excitation, or more often, a whole series of excitations, belongs to the most basic tools in Landau level spectroscopy. When we follow this dependence, we can quickly assess the shape of the bands responsible for the observed response.

To illustrate how this assessment typically works, let us look at a recent example of Landau level spectroscopy performed on ZrSiS~\cite{ShaoPRX24}. This material has a fairly complex band structure \cite{Schilling2017,UykurPRR19} that leads to a rich magneto-optical response shown in Fig.~\ref{fig:Semidirac}a. Three series of inter-Landau-level excitations have been identified and indicated by dashed and colored lines. Each series has a characteristic dependence on the applied magnetic field. 
At high energies, roughly above 300~cm$^{-1}$, we can see a typical signature of massive Dirac electrons: transitions with a sub-linear dependence in $B$ which in the limit of a vanishing magnetic field extrapolate to the local band gap of $2\Delta$. 
Similar transitions have previously been observed in many other materials \cite{OrlitaPRL15,AssafSR16,AssafQM17,JiangPRB17,Le-Mardele:2023,MohelskyPRB26}.
The remaining two series of transitions both originate in the gapless parts of the band structure, but they evolve differently as a function of the applied magnetic field. While the series with the $\sqrt{B}$ dependence comes from a conical band~\cite{SadowskiPRL06,JiangPRL07,DeaconPRB07, OrlitaPRL08II,BordacsPRL13, Orlita2014}, the $B^{2/3}$ dependence is interpreted in terms of semi-Dirac electrons with an anisotropic dispersion. Depending on the direction, this dispersion can be linear or quadratic. The task of further investigation, supported by {\em ab initio} methods, then becomes to assign these three different series to particular locations in the bulk  or the surface band structure of ZrSiS.

\begin{figure}[!th]
	\includegraphics[width=\linewidth]{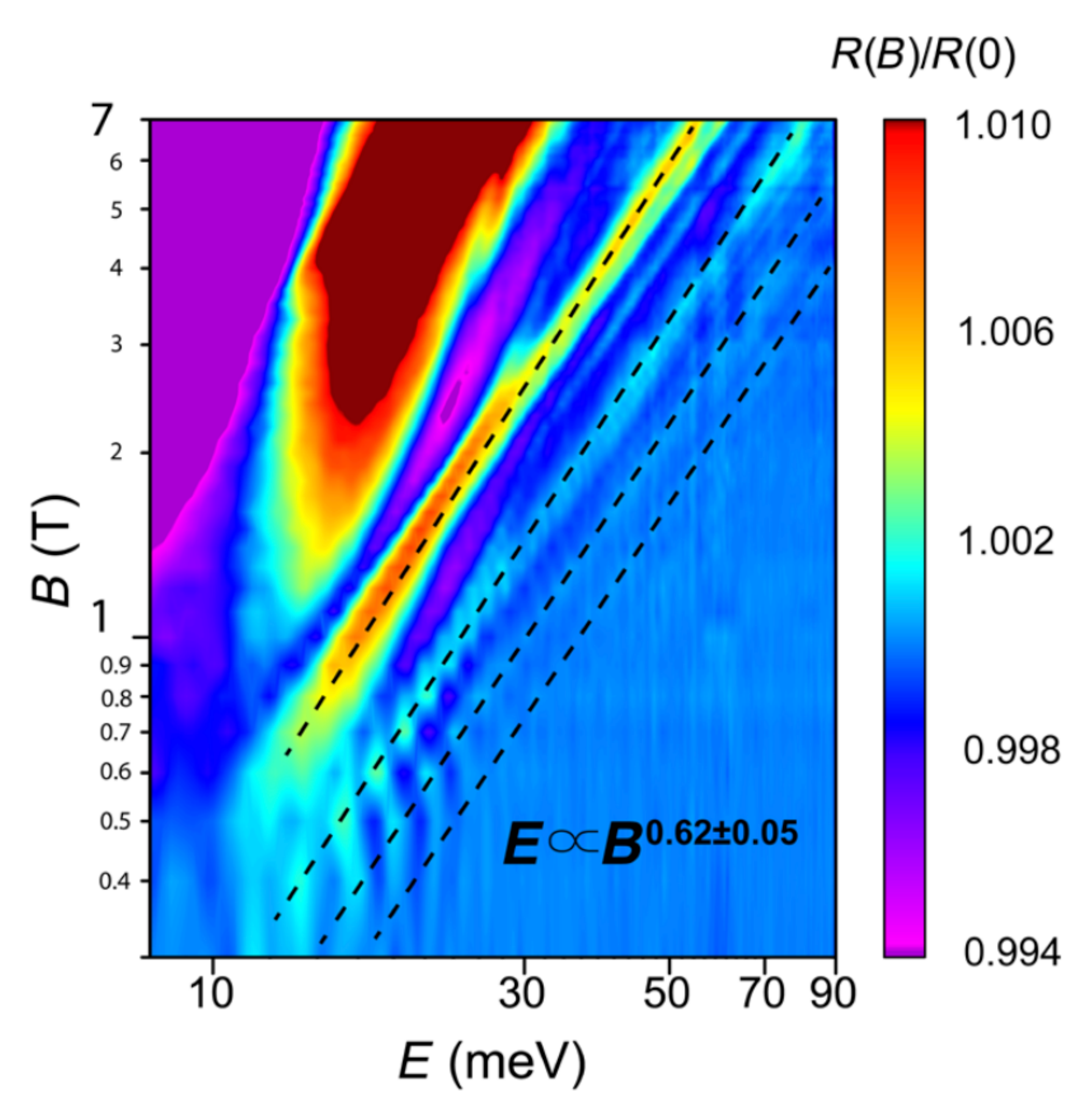}
	\caption{    
    Color plot of the normalized reflectance spectrum of TaAs, measured in Voigt geometry. The dominant inter-Landau-level transitions can be described phenomenologically by a scaling of $B^{0.62}$, as indicated by dashed lines. The deviation of the scaling from $B^{0.5}$, expected for perfectly linearly-dispersing bands near the Weyl points, was understood in terms of the band curvature near the saddle point that connects the two Weyl points. Adapted from \cite{LuSA2022}.}
	\label{fig:TaAs}
\end{figure}

Inter-Landau-level transitions that are perfectly linear in $B$ or $\sqrt{B}$ have been observed in the magneto-optical response of a number of materials. In such materials, Landau level spectroscopy has proven the presence of an ideal parabolic or conical band, within the experimental accuracy set by the broadening of resonances, thereby reflecting the electronic quality of explored crystals. However, in many other cases, these simple scaling laws can only serve as a rule of thumb. One reason is that the presence of the periodic band structure in crystalline solids implies that any parabolic or conical band eventually flattens at sufficiently large momentum. Resulting departures from the ideal parabolic or conical shape can be traced using Landau level spectroscopy. The approximate validity of the scaling laws is illustrated in the example of TaAs, shown in Fig.~\ref{fig:TaAs}. This material was among the first experimentally confirmed Weyl semimetals~\cite{LvPRX15,ArmitageRMP18}, but the observed response deviates from that of an ideal Weyl semimetal. Its dispersion is anisotropic and does not follow the perfect $\sqrt{B}$ dependence for all directions of the applied magnetic field~\cite{LuSA2022,Santos-CottinPRB22}.

\begin{figure}
    \centering
    \includegraphics[width=0.95\linewidth]{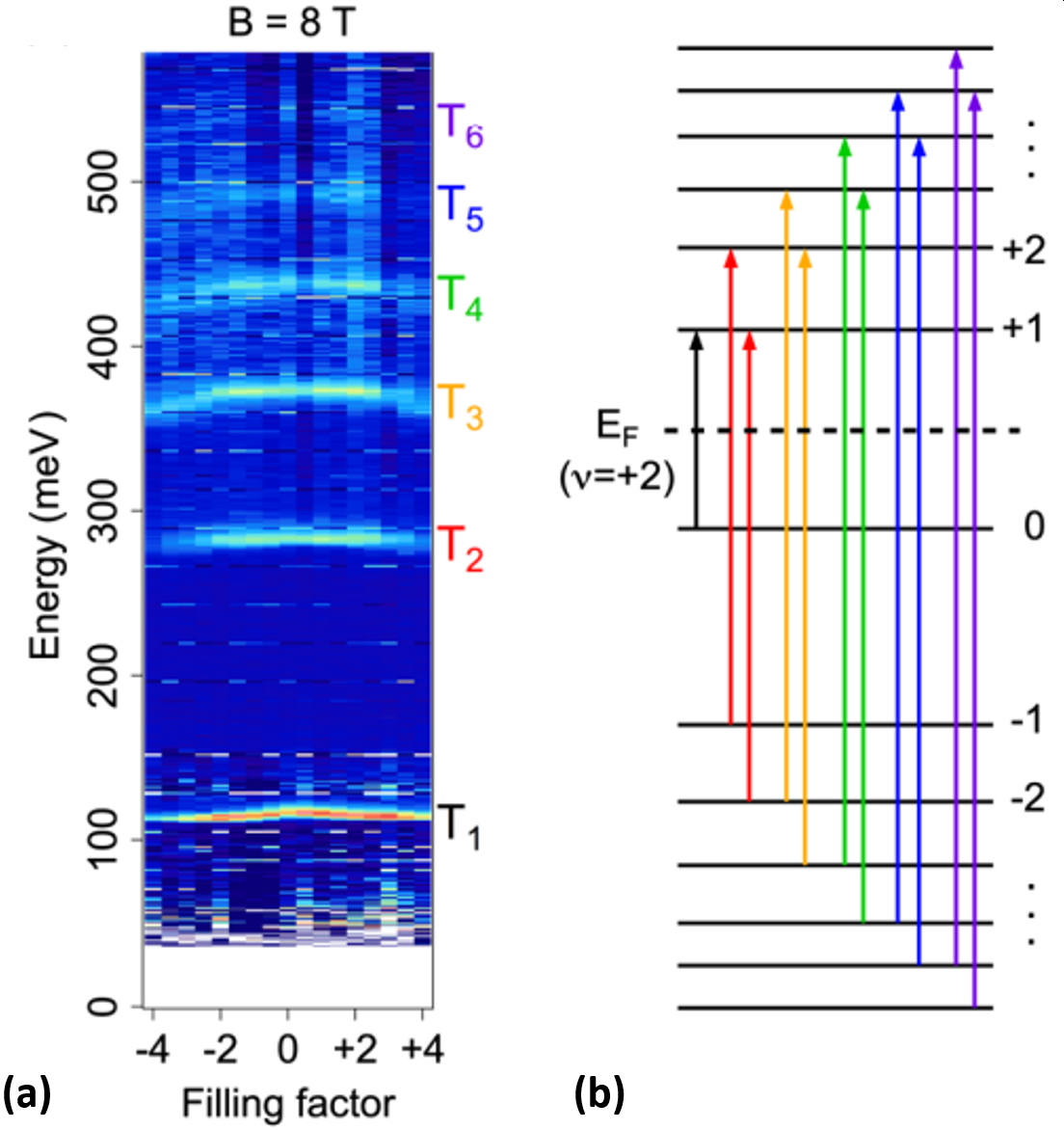}
    \caption{False-color map of graphene magneto-absorbance as a function of carrier density (filling factor $\nu$) accompanied by a schematic diagram indicating the resonances observed at $\nu=+2$. Evidently, the transition energies $T_n$ depend on the filling factor. Reprinted with permission from~\cite{RussellPRL18}, copyright (2018) by the American Physical Society.}
    \label{fig:Henriksen}
\end{figure}

In the above examples, we used the single-particle approach, in accordance with the Kubo-Greenwood formula~\ref{Kubo}. In other words, we assumed that the energies of inter-Landau-level excitations exactly match the distance between pairs of Landau levels. Strictly speaking, this is valid only for cyclotron resonance in materials with a perfectly parabolic band. 
There, Kohn's theorem holds~\cite{KohnPR61}, and the electron-electron interaction does not change the cyclotron resonance frequency. In all other cases, the Coulomb interaction among electrons impacts the observed response. The single-particle model usually remains remarkably valid, but only as an approximation. 
This can be nicely seen in the example of interband excitations in semiconductors with a wider band gap, where excitonic effects play an important role, particularly close to the interband absorption edge~\cite{ZwerdlingPR59,VrehenJPCS68}. Another excellent example is presented in Fig.~\ref{fig:Henriksen}, which shows magneto-absorbance in a high-quality graphene monolayer, as a function of the carrier density described by the filling factor $\nu$. The experiments showed a series of inter-band inter-Landau-level resonances typical of graphene~\cite{RussellPRL18}. However, not only their intensity and broadening, but also their resonance energies evolve with the carrier density. This means that the single-particle approach is still applicable, but only approximately. 

\begin{figure*}[t]
    \centering
    \includegraphics[width=0.8\linewidth]{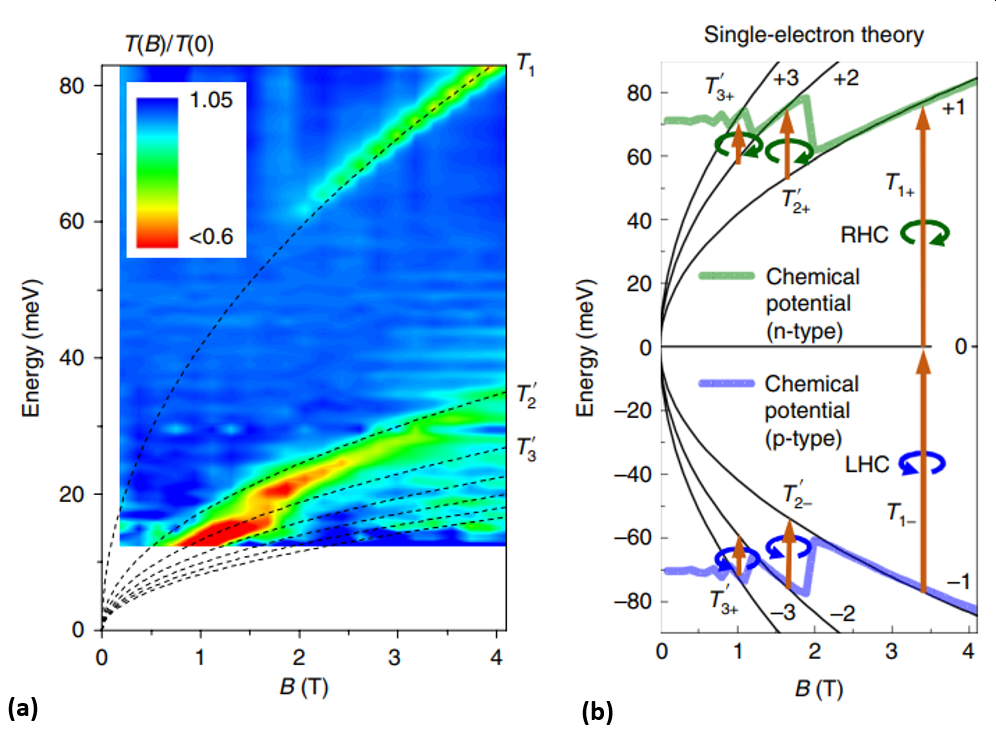}
    \caption{(a) Color map of magneto-transmission in h-BN encapsulated, weakly doped graphene measured with unpolarized radiation. 
    (b) Blue and green lines show the calculated magnetic field dependence of the chemical potential for $n$-type and $p$-type doping.
    Adapted from Ref.~\cite{NedoliukNN19}.}
    \label{Fig:Nedoliuk}
\end{figure*}

If we consider Eq.~(\ref{Kubo}) more realistically, we must take into account that all solids have both extrinsic and intrinsic disorder. The infinitely sharp resonance then gains a finite linewidth. Typically, the $\delta$-function profile is replaced by a Lorentzian: $\delta(E_m-E_n-\hbar\omega)\rightarrow \frac{1}{\pi}\frac{\Gamma}{(E_m-E_n-\hbar\omega)^2+\Gamma^2}$, where $2\Gamma$ stands for the full width at half maximum (FWHM) of the absorption line. Alternatively, a Gaussian or a mixed Gaussian-Lorentzian profile may appear in systems with inhomogeneous broadening. In three-dimensional materials, the line broadening can also have an intrinsic origin, caused by the electron dispersion along the magnetic field direction~\cite{OrlitaPRL08,Orlita2014}. This may result in the asymmetry of the resonances, which then gain pronounced high- or low-energy tails.

The line broadening and its dependence on temperature, energy and magnetic field can give important insights into electronic scattering mechanisms. A large body of experimental and theoretical knowledge connects different types of scattering mechanisms (such as long- or short-range neutral impurities, charged impurities, and so on) to different kinds of Landau level broadening, and also directly to the cyclotron resonance absorption width. The ways in which the cyclotron resonance width depends on the applied  magnetic field and temperature are discussed extensively, for example, in \cite{AndoJPSJ75,EnglertSSC83,SchlesingerPRB84,ShonJPSJ98,YangPRB10}. In comparison, studies on how interband inter-Landau-level resonances broaden because of disorder are not as conclusive. Still, there are certain empirical rules, such as the tendency of Landau levels to broaden with increasing index $n$, magnetic field, or energy. Such behavior is observed and discussed, e.g., in \cite{OrlitaPRL11,NedoliukNN19,TikuisisPRB21}.

\subsection*{Occupation effect} 

The occupation factor $f_n-f_m$ represents the difference between the occupation of the initial and final Landau levels, denoted by $n$ and $m$. In the limit of low temperatures---when the Fermi-Dirac distribution takes the form of a step-like function---this factor reaches a binary value, 0 or 1. At higher temperatures, when the separation  of Landau levels $E_m-E_n$ becomes comparable to or smaller than $k_B T$, the occupation factor weighs the contribution of electron excitations between different pairs of Landau levels. The occupation factor often allows us to read out the Fermi energy or the carrier density directly from the data, without the need for any transport measurements \cite{LuSA2022,Akrap2016,Martino2019}.

Cyclotron resonance absorption in graphene is a particularly clear example, as we can see in magneto-transmission experiments shown in Fig.~\ref{Fig:Nedoliuk}a. The occupation of Landau levels, and the corresponding occupation factor $f_n-f_m$, is determined by the Fermi-Dirac distribution, the number of electrons, and the Landau level degeneracy ($\zeta=eB/h$). For instance, when the $0 \rightarrow 1$ transition emerges in the spectrum around $B=2$~T, we know that the occupation $f_1$ of the $n=1$ Landau level is slightly below 1. Hence, at $B=2$~T, the number of occupied Landau levels (the so-called filling factor)  reaches $\nu=6$, taking into account the double spin and valley degeneracy in graphene, as well as the specific nature of the $n=0$ Landau level, shared between the conduction and valence bands. From these values we can directly calculate the electron density, $N=\nu\cdot\zeta\approx 3\times10^{11}$~cm$^{-2}$. 

\subsection*{Selection rules} 
The factor ${\cal{M}}_{m,n}$ describes the probability that an electron in the $n$-th Landau level will be promoted by light into the $m$-th Landau level. This probability reflects the properties of the explored system, encoded in the Hamiltonian. It also depends on the polarization state of light (circular versus linear) and the propagation direction of light with respect to the applied magnetic field. Formally, this factor is expressed through the matrix elements of the interaction Hamiltonian: ${\cal{M}}_{m,n}=|\left <m|H_{\mathrm{int}}|n\right>|^2$. In the majority of cases relevant for Landau level spectroscopy, these matrix elements are calculated within the electric-dipole approximation, but the form of expression (\ref{Kubo}) does not exclude other, usually weaker absorption lines, such as magnetic-dipole or electric quadrupole transitions.

\begin{figure*}
    \centering
    \includegraphics[width=0.8\linewidth]{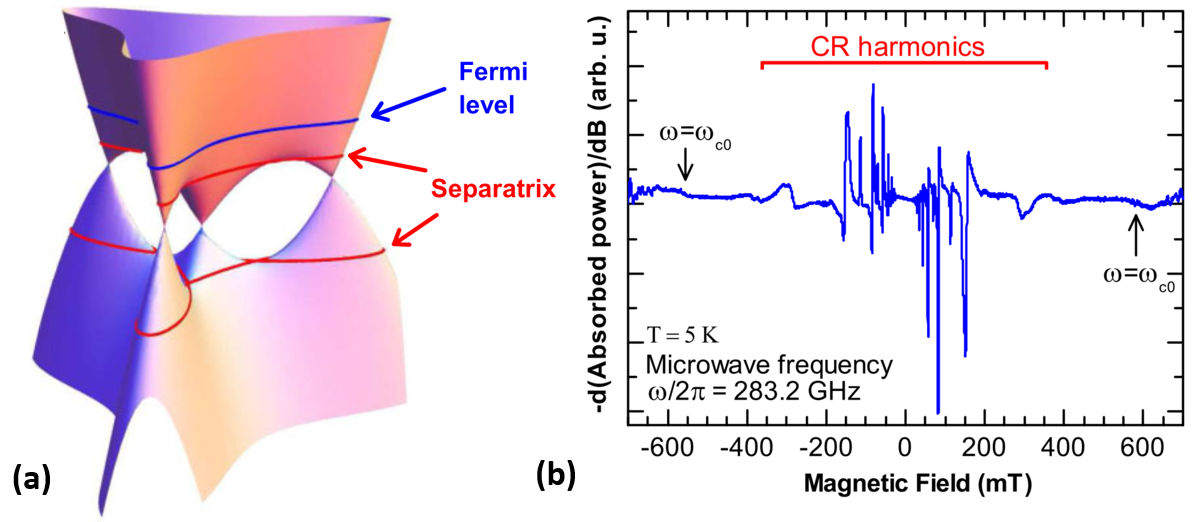}
    \caption{(a) Electronic band structure of graphite at the $K$ point. The Fermi level is located just above the Lifshitz transition and it is its shape is highly impacted by the triangular symmetry. (b)
    Cyclotron resonance caused by electrons in trigonally warped bands at the $K$ point with a number of harmonics. Reprinted with permission from~\cite{Orlita2012}, copyright (2012) by the American Physical Society.}
    \label{fig:graphite}
\end{figure*}

Two experimental arrangements are often employed in Landau level spectroscopy. One has the light propagating along, and the other perpendicular to the applied magnetic field. The two configurations are called the Faraday and Voigt configurations. In the Faraday geometry, the symmetry implies that light has two circularly polarized  eigenmodes. Typically, this gives rise to two series of excitations: $m=n + 1$ and $m=n - 1$, both of which preserve the spin orientation. In a simple interpretation, when one circularly polarized photon is absorbed, the quantized angular momentum of an electron undergoing cyclotron motion is raised or lowered by a quantum of angular momentum $\hbar$. For intraband resonances (cyclotron resonance absorption), the above selection rules imply that only transitions between adjacent levels are possible. In parabolic bands, where adjacent Landau levels are separated by the classically defined cyclotron frequency $\omega_c=eB/m$, see Eq.~(\ref{Schrodinger}), both quasi-classical and fully quantum mechanical treatment lead to the same result. 

In the Voigt geometry, the two eigenmodes of light are linearly polarized --- and this can be along or perpendicular to the applied magnetic field. In the latter case, the selection rules remain the same as in the Faraday configuration, $m=n \pm 1$. In the former case, the selection rule gives rise to interband inter-Landau-level excitations that preserve the index, $n=m$. In addition, $n=m$ transitions accompanied by a spin-flip may emerge, with an intensity that scales with the strength of the spin-orbit coupling. 
The magneto-reflectivity data presented in Fig.~\ref{fig:TaAs}, collected on TaAs using non-polarized light, is an example of a Landau level spectroscopy measurement realized in the Voigt configuration. Both series of transitions, $m=n \pm 1$ and $m=n$, are found in the response, see \cite{LuSA2022} for details.

Nevertheless, we should not be surprised that some of the selection rules in the literature can be different or considerably more complex than those mentioned above. The reasons for this can vary. The selection rules naturally depend on the way in which individual Landau levels are indexed or numbered. Studies of Landau-quantized zinc-blende semiconductors may serve as a good example. There, the literature survey may suggest that the selection rules for interband excitations between Landau levels are $n= m $ and $n = m-2$  \cite{PidgeonPR66}, while in other works the selection rules read $n = m \pm 1$ \cite{WeilerSaS81}. In reality, different theoretical approaches to calculate Landau levels lead to different selection rules, but both describe the same set of experimentally observed inter-Landau-level excitations. The selection rules therefore cannot be taken separately from the theoretical model of Landau quantization.

Apart from this formal reason, there are sometimes specific physical reasons to observe transitions beyond the simple selection rules sketched above. This happens, for instance, when the symmetry imposed by the experimental configuration is further reduced by the symmetry of the explored material. We may illustrate this with an example of electrons at the $K$ point of bulk graphite, where the conduction band has profound trigonal warping as illustrated in Fig.~\ref{fig:graphite}a. In a simple parabolic band, we expect a single cyclotron resonance to be active in only one circular polarization, which depends on whether we are dealing with electrons or holes. In graphite, however, we observe a series of excitations, shown in Fig.~\ref{fig:graphite}b, which are active in both circular polarizations. The basic cyclotron resonance mode at the frequency $\omega_{c0}$ is accompanied by a series of lines with the selection rules: $m=n + 1+3j$ and $m=n - 1+3j$ ($j=0,1,2,\ldots$)~\cite{Orlita2012}. Hence, when the full rotation symmetry along the applied magnetic field is reduced, in this case to a trigonal symmetry, the cyclotron resonance harmonics become activated. This is why we can see resonances at frequencies that are selected multiples of $\omega_{c0}$. 

These unexpectedly rich selection rules can still be understood in terms of classical cyclotron motion. In a warped electronic band, the cyclotron motion of an electron remains periodic with a fundamental frequency of $\omega_{c0}$. However, because the cyclotron trajectory is strongly curved, its motion necessarily contains higher harmonics. These higher harmonics can also couple to electromagnetic radiation. Moreover, the curvature of the trajectory changes sign six times during a single cyclotron orbit. Because of this, both left- and right-circularly polarized light become cyclotron-resonance-active in this specific case.

\begin{figure}[th]
	\includegraphics[width=\linewidth]{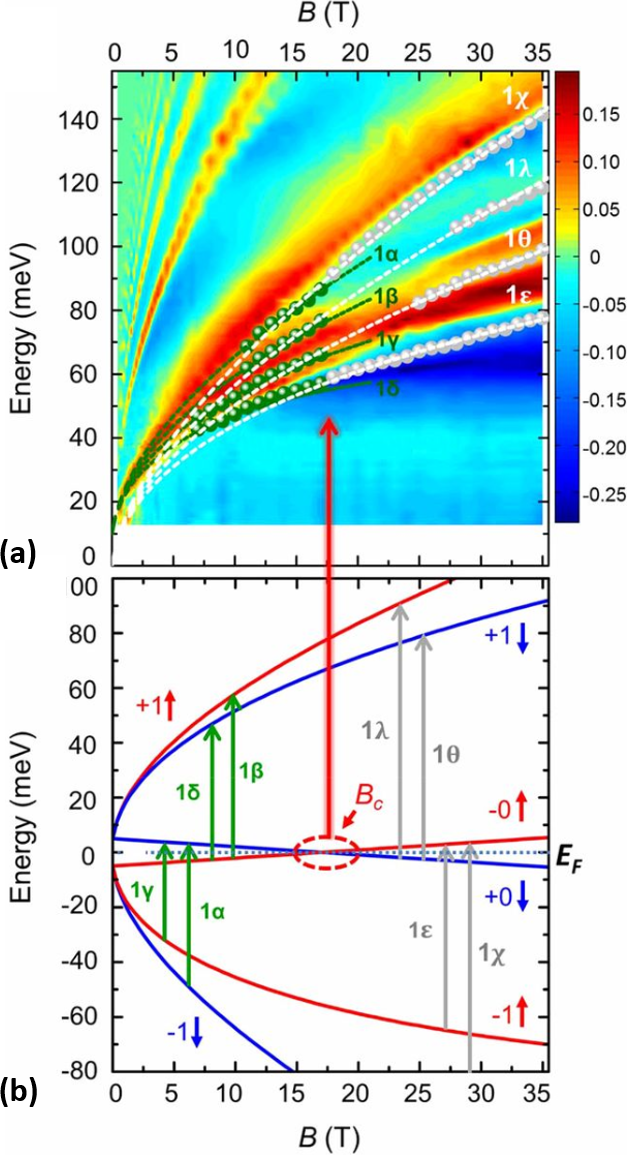}
	\caption{    
    (a) Color plot of the normalized absorbace spectrum of ZrTe$_5$ as a function of magnetic field and energy. Quartets of transitions, identified in (b) by vertical arrows, are shown in green and white spheres. This is a consequence of a strong spin-orbit coupling, leading to Zeeman splitting.
    Adapted from \cite{Chen2017}.}
	\label{fig:ZrTe5}
\end{figure}

Yet another illustration of selection rules that go beyond basic expectations is presented in Fig.~\ref{fig:ZrTe5}. Magneto-transmission measured in the Faraday configuration on ZrTe$_5$---a semiconductor with a nearly vanishing band gap and, therefore, almost conical bands---shows a response richer than anticipated in the quantum limit, when the Fermi level is confined within the $n=0$ Landau levels. Instead of a doublet, expected within the basic $0\rightarrow 1$ and $-1\rightarrow 0$ selection rule, a quartet is observed. This is due to Zeeman splitting of levels and additional selection rules that allow not only spin-conserving, but also spin-flipping transitions to appear. The additional selection rules are connected with the overall crystalline symmetry of ZrTe$_5$ and the strong spin-orbit coupling in this material.

\section{Landau level spectroscopy -- optical techniques}

\begin{figure}[th]
	\includegraphics[width=\linewidth]{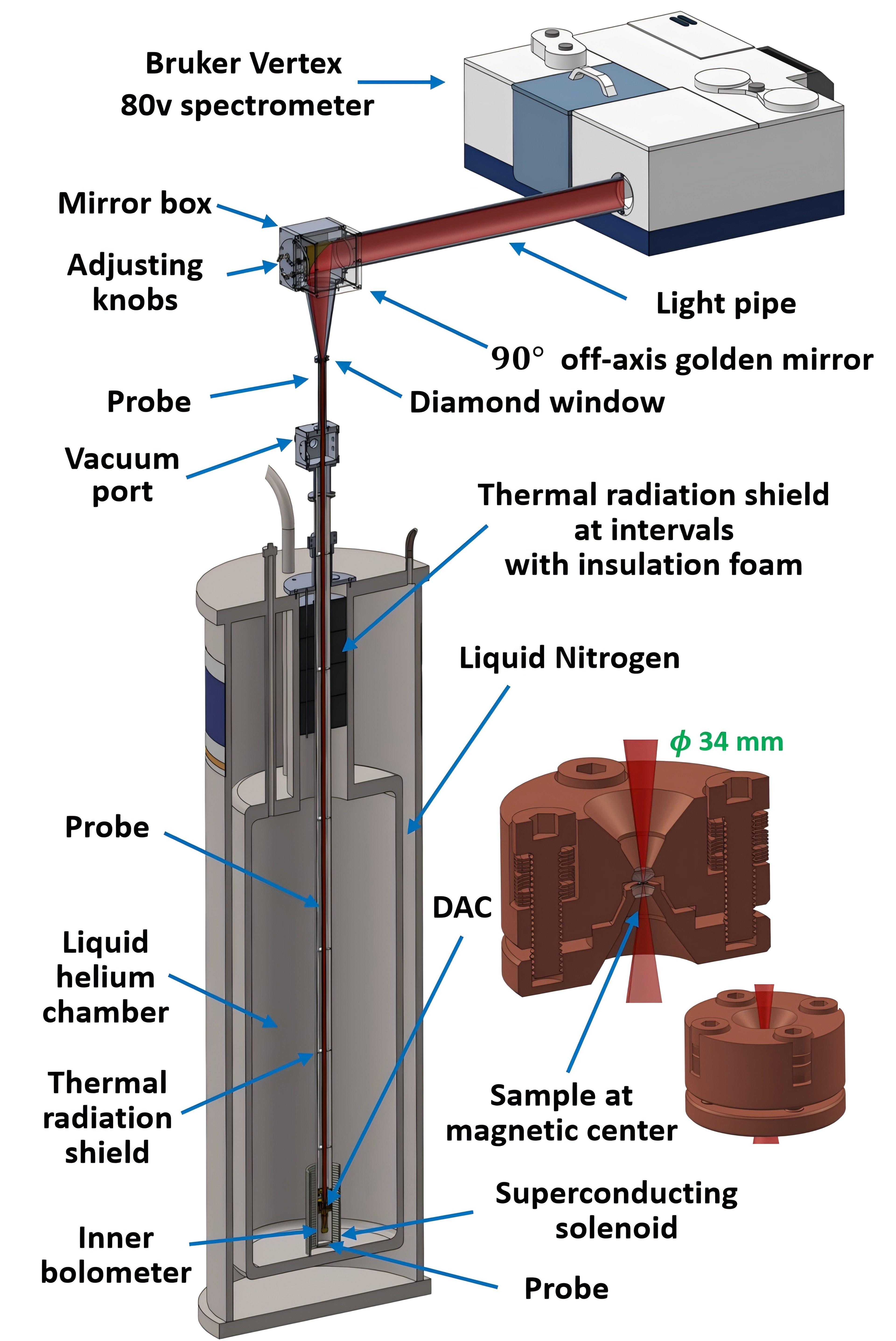}
	\caption{
    Illustration of a typical THz/infrared spectroscopy setup for low-temperature measurements in strong static magnetic fields. Radiation from a thermal source (typically globar or mercury arc lamp) is modulated interferometrically inside the Fourier transform spectrometer and delivered, using light-pipe and reflective optics, to the sample kept at low temperatures in the superconducting coil. The radiation is then detected by the bolometer placed directly below the sample. In this particular case, the sample is located in a diamond anvil cell (DAC), which is illustrated in the lower right corner. Adapted from \cite{LvRSI25}.}
	\label{fig:setup}
\end{figure}

Various optical techniques have been coupled with magnets to realize Landau level spectroscopy experiments. The utilized experimental methods span spectrally from the sub-THz to UV range, and the applied magnetic fields vary by several orders of magnitude, from milliteslas~\cite{NeugebauerPRL09} to several hundred of teslas~\cite{Kono02b}. The choice of a specific technique depends on the nature of the system explored: metals require experimental conditions different from those of semimetals and semiconductors. The experimental setup is also chosen following the type of inter-Landau-level excitations that one plans to study, whether these are intraband excitations (cyclotron resonance) or interband excitations.

Historically, most commonly used optical experiments had fixed monochromatic sources of radiation and tracked changes in reflectivity and transmission as a function of the applied magnetic field. This was how the first cyclotron resonance measurements were performed in the early 1950s~\cite{Dresselhaus:1953,CohenAIPCP2005}.
Currently, this approach is used for fast Landau level spectroscopy measurements in pulsed magnetic fields~\cite{Kono02b,NicholasPRL13,MiyataPRB17}. 
In parallel, experimental techniques using broadband radiation and grating-type spectrometers have been developed in 1950s and regularly employed, see Fig.~\ref{fig:EarlySetup}, yet encountering a number of experimental challenges when approaching the long-wavelength optical limit: down to the far infrared and THz spectral ranges.

The real breakthrough happened when interferometric techniques were introduced and Fourier transform spectrometers became commercially available ~\cite{KaplanAO76,Griffiths06}. This enabled probing Landau-quantized solids effectively, in a broad range of photon energies, down to sub-THz frequencies. 
Figure \ref{fig:setup} shows a typical current setup for Landau level spectroscopy. It incorporates a modern Fourier transform infrared spectrometer for measurements in magnetic fields produced by a superconducting coil. The Fourier transform technique not only made it possible to measure the absolute reflectivity and transmission in high-field experiments, allowing one to reliably determine the resonance line shapes, but it also paved the way towards the full Kramers-Kronig-type analysis of the optical data collected on Landau-quantized solids, with the real and imaginary components of the response function extracted~\cite{LevalloisRSI15}. 

Today, Landau level spectroscopy is a well-established optical approach to probe Landau-quantized solids, using various spectroscopic techniques coupled to magnetic field sources. These techniques include Fourier transform spectroscopy, ellipsometry~\cite{KuhnePRL13}, electron paramagnetic resonance (EPR) spectrometers~\cite{NeugebauerPRL09,Orlita2012}, Raman spectroscopy~\cite{YanPRL10,KossackiPRB11,FaugerasJRS18}, conventional grating-type spectrometers and, more recently, also time-domain THz spectroscopy~\cite{ArikawaPRB11,ZhangPRL14,RieplAPL25}.  Elaborate experimental setups allow for measurements under pressure \cite{LvRSI25}, as well as measurements with spatial resolution far beyond the diffraction limit~\cite{WehmeierSA24}.

\section{Landau level spectroscopy -- perspectives}

Landau level spectroscopy is an established technique which will certainly remain in the portfolio of basic experimental methods to investigate and understand crystalline solids. It will continue to be a primary way to get precise estimates of effective masses and band gaps in emerging two-dimensional and thee-dimensional materials. We envisage that among those new materials there will be magnetic topological systems, including dilute magnetic topological insulators \cite{Tokura:2019_NRP}, and  strongly correlated topological materials.

In the coming years, we may expect a push towards Landau level spectroscopy measurements with a spatial resolution going beyond the diffraction limit. This development is motivated mainly by the increasingly important area of two-dimensional materials (i.e., atomic monolayers, homo- and hetero-bilayers, with or without moire patterns). These materials are often available only as crystallites whose lateral dimensions do not exceed several microns. At present, such infrared/THz magneto-spectroscopy measurements are only possible using a scanning near-field optical microscopy (SNOM) technique, when adapted to operate at low temperatures and in magnetic fields, see, e.g., Refs.~\cite{KimRSI23,WehmeierSA24}. 

Magneto-Raman scattering is another way to bypass the diffraction limit of THz/infrared spectroscopy. One may study the low-energy excitations with energies in the THz and infrared range, but with a micrometer spatial resolution, given by the diffraction limit in the visible spectral range. Experiments performed on few-micron-sized graphene flakes~\cite{BerciaudNL14,NeumannNL15,FaugerasJRS18} demonstrate the feasibility of this method in Landau-level spectroscopy of 2D materials. 

Experiments that follow the dynamics of Landau-quantized photo-excited carriers are another emerging direction where we can expect important developments. The recent progress in generating intense pulsed and THz radiation, mastered at free-electron-laser facilities, allows for such experiments in magnetic fields that reach several teslas. These experiments are typically performed on well characterized materials. The goal is no longer to gain an elementary understanding of the material in question. Instead, the primary focus becomes understanding the relaxation mechanisms and interaction of electrons with other quasi-particles, in particular with other electrons ~\cite{MittendorffNP15,Konig-OttoNL17,ButNPhot19}. Alternatively, time-resolved cyclotron measurements can also be performed due to recent progress in time-domain THz spectroscopy~\cite{ArikawaPRB11,ZhangPRL14,GeorgeJOSAB12,RieplAPL25}.

One may also anticipate that Landau level spectroscopy based on other, non-optical techniques, will progress as well. At the moment, the STM/STS at low temperatures and in magnetic fields appears to be a powerful method. Its applicability has already been demonstrated in a number of experimental examples, including graphene, graphite, 3D Dirac semimetals, or surface states of topological insulators~\cite{LiNP07,LiPRL09,HanaguriPRB10,LiuNM14}. Even though there are significant experimental constraints, angle-resolved photoemission (ARPES) on samples subjected to a magnetic field may be a viable future development. At present, the magnetic field available experimentally remains fairly weak~\cite{RyuJESRP23}, measured in tens of militesla. While further progress is desirable and anticipated, the magnetic field available now is already capable of inducing Landau quantization in certain solids, notably in graphene~\cite{NeugebauerPRL09,MayorovNC26}.

\section{Concluding remarks}

This review introduces the fundamental aspects of Landau level spectroscopy and places it into the historical context of important discoveries in solid state physics. We discuss the concept of cyclotron motion of electrons in solids and related resonant absorption of light. In magnetic fields that are high enough, this gives way to a quantization in Landau levels and a possibility to promote electrons among them by means of light. The optical response of a Landau-quantized solid is dissected and illustrated on selected examples taken from the literature, principally on graphene and topological materials.

Our educational review does not address more than truly the basics of Landau level spectroscopy. For a deeper immersion into the topic, several books and reviews can be recommended. Some of them have already been cited in the text. First, there is a classical solid-state book \emph{Landau level spectroscopy} edited by G. Landwehr and E. I. Rashba, which represents the state-of-the-art in this particular field in the early nineties. The basics of Landau quantization are introduced in well-known textbooks, such as the ones by N. Ashcroft \& N.~D. Mermin~\cite{AshcroftMermin1976}, C. Kittel \cite{Kittel:1954} or J. Singleton \cite{Singleton01}. The cyclotron resonance response in solids has been described in depth, for example, by Palik and Furdyna~\cite{PalikRPP70}, by T. Ando et al.~\cite{AndoRMP82}, and later on also by J.~Kono and N.~Miura~\cite{Kono02a,Kono02b}. 
For more recent developments, several books and review articles are available ~\cite{Noboru07,GoerbigRMP11,GornikNPhot21}.

Landau level spectroscopy is an established technique with a long and distinguished past. Currently, it is being actively used by a handful of experimental solid-state physics groups around the world. Landau level spectroscopy provides us with a wealth of information about semiconductors and semimetals, be they topological or not, and will for sure continue to do so in the years to come.

\section*{Acknowledgments}
We are grateful for discussions with M. Potemski and assistance by F. Le Mardel\'e. M.O. acknowledges the support by the ANR, via project TEASER (ANR-24-CE24-4830). A.A. acknowledges the support of the Croatian Science Foundation under Project No. IP-2025-02-1189.
This article is dedicated to the memory of Mark O. Goerbig, who taught us about Landau levels, graphene, wine and friendship.


%

\end{document}